\documentclass[aps,floats,epsf,twocolumn]{revtex4}

\newcommand{\baromega}{{\bar{\omega}}}

\usepackage{graphicx}
\usepackage{amsmath,amssymb}
\usepackage{bm}

\begin{document}

\title{Skyrmion in spinor condensates and its stability in trap potentials}

\author{A. Tokuno$^{1}$}
\author{Y. Mitamura$^{2,3}$}
\author{M. Oshikawa$^{2}$}
\author{I. F. Herbut$^4$}
\affiliation{$^1$ Department of Applied Physics, Hokkaido University,
Sapporo 060-8628, Japan}
\affiliation{$^2$ Institute for Solid State Physics, University of Tokyo,
Kashiwa 227-8581, Japan}
\affiliation{$^3$ Department of Physics, Tokyo Institute of Technology,
Oh-okayama, Meguro-ku, Tokyo 152-8551, Japan }
\affiliation{$^4$ Department of Physics, Simon Fraser University,
Burnaby, British Columbia, Canada V5A 1S6}

\date{\today}

\begin{abstract}
 A necessary condition for the existence of a skyrmion in two-component
 Bose-Einstein condensates with $\mathrm{SU(2)}$ symmetry
 was recently provided by two of the authors
 [Phys. Rev. Lett. {\bf 97}, 080403 (2006)],
 by mapping the problem to a classical particle
 in a potential subject to time-dependent dissipation.
 Here we further elaborate this approach.
 For two classes of models, we demonstrate the existence of
 the critical dissipation strength above which the skyrmion
 solution does not exist.
 Furthermore, we discuss the local stability of the skyrmion
 solution by considering the second-order variation.
 A sufficient condition for the local stability is given
 in terms of the ground-state energy of a one-dimensional
 quantum-mechanical Hamiltonian.
 This condition requires
 a minimum number of bosons, for
 a certain class of the trap potential.
 In the optimal case, the minimum number of bosons can be as
 small $\sim 10^4$.
\end{abstract}
\maketitle

\section{Introduction}

  Topological defects often play a fundamental role in our understanding of
phases of matter and the transitions between them. The best understood
examples are probably vortices and vortex-loops in superfluids and
$\mathrm{O(2)}$ magnets in two and three dimensions, which are
responsible for the very existence of the high-temperature phase, and
completely determine the universality class of the phase
transition~\cite{herbutbook}. The next in order of complexity are the
topological defects in the $\mathrm{O(3)}$-symmetric Heisenberg model,
which allows skyrmions in two dimensions~\cite{belavin} and
hedgehogs-like configurations (a point defect around which spins point
outward) in three dimensions~\cite{MS}. While the former only
renormalize the coupling constant, the role of the latter is less
clear~\cite{dasgupta}. All of the above, however, represent 
configurations topologically distinct from vacuum, which provides them
with local stability.

  In this paper, we study the stability of the topologically non-trivial
skyrmion configuration in  theories with $\mathrm{O(4)}$ symmetry. Such a
symmetry arises in complex condensates with an internal
spin-$1/2$-like quantum number~\cite{lieb}, for example. Realizations
of such {\it spinor condensates} are found in models of inflatory
cosmology~\cite{vilenkin}, Bose-Einstein condensation of
$^{87}$Rb~\cite{ueda}, and bosonic ferromagnetism~\cite{lieb,saiga}, and
in effective theories of high-temperature
superconductivity~\cite{herbut,zlatko}~and of deconfined
criticality~\cite{senthil}.  The Higgs sector of the Weinberg-Salam
model of electroweak interactions represents another closely related
example, with a spinor condensate coupled to gauge fields.
In our problem in three dimensions, there exists a topologically
non-trivial mapping to the order-parameter space, thanks to the fact
that the {\it third} homotopy group of $S^3$ is the  group of integers.
However, the topology alone turns out to be insufficient to guarantee
local stability of the skyrmion. This may be understood already in terms
of the classic Derick theorem~\cite{hobart}. Recently, a more general
proof that the skyrmion cannot be a stationary point of the action for
the spinor Bose-Einstein condensate (BEC) in free space was given by two
of the present authors~\cite{HO}, based on the analogy between the
Euler-Lagrange equations and the classical mechanics of a particle in a
time-dependent dissipative environment~\cite{durrer}. The advantage of
this alternative point of view at the old problem is that it provides one
with a simple way of constructing the external potentials which would
indeed lead to skyrmion as the solution of the Euler-Lagrange equations.
In Ref.~\cite{HO}, three such special potentials were presented.
In this paper, we further develop this approach.

First we analyze the construction of the solution based on
the ansatz proposed in Ref.~\cite{HO}.
There, a time-dependent dissipation which is odd in time
was introduced, to allow an odd solution.
However, we find that the symmetry argument does not always work,
and there is a critical dissipation strength above which
the odd solution no longer exists even for an odd dissipation.
Next, we study the stability of the skyrmion in a generalized class of
the potentials introduced in~\cite{HO} with respect to small
variations. We map the problem to an effective quantum-mechanical
eigenvalue problem and determine the region of local stability in the
parameter space. A particularly interesting result of our analysis is
that the stable skyrmion requires a minimal number of particles in the
trap, estimated here to be $\sim 10^4$ in the optimal case.

The paper is organized as follows. In Sec.~\ref{skyrmion_solution}, we
present basic formulation of the problem. A classical equation of motion
with time-dependent dissipation determines the skyrmion solution. On the
other hand, a quantum mechanical eigenvalue problem determines the local
stability of the solution. In Sec.~\ref{construction}, we discuss the
construction of the skyrmion solution based on the odd-function ansatz.
We determine the critical dissipation strength, which separates the
region with and without a skyrmion solution. In Sec.~\ref{numerical}, we
present the numerical solution of the equation of motion for a few
cases. The existence of the critical dissipation strength, as well as
related theoretical predictions, is confirmed numerically.
Furthermore, the local stability of the obtained solution is
also analyzed by solving the quantum-mechanical eigenvalue
problem.
The minimal number of bosons required to satisfy the sufficient
condition for the local stability is numerically obtained as a function
of parameter. Section~\ref{summary} is devoted to summary of the paper.

\section{General discussion on the skyrmion solution}
\label{skyrmion_solution}

\subsection{Basic equations}

We begin by reviewing the derivation of the skyrmion solution for the
two-component (spinor) BEC in an external potential, formulated
previously in Ref.~\cite{HO}.
The derivation is based on the Euler-Lagrange equations, and leads to
the necessary condition for the existence of the skyrmion solution.
We will then proceed to examine the stability of the skyrmion by taking
into account the second-order variation.

Let us consider the two-component bosons in three-dimensional continuum
space in the  external confinement potential $V(\bm{r})$. The system can
be described by the following effective action via the path integral
formalism~\cite{herbutbook}:
\begin{align}
 S
 &=\int_{0}^{\beta}\!\!\! d\tau
   \!\!\int\!\!d^3\bm{r}\
   {\cal L}
 \nonumber \\
 {\cal L}
 &=\Phi^{\dagger}
   \left(
    \partial_{\tau}
    -\frac{\hbar^2}{2m}\nabla^2
    -\mu
    +V(\bm{r})
   \right)
   \Phi
   +\frac{U}{2}
    \left(
     \Phi^{\dagger} \Phi
    \right)^2,
 \label{action}
\end{align}
where
$\Phi^{\dagger}(\tau,\bm{r})=(\Phi_{1}^{*}(\tau,\bm{r}),\Phi_{2}^{*}(\tau,\bm{r}))$
is a two-component bosonic field with the mass $m$, which satisfies the
periodic boundary condition $\Phi(0,\bm{r})=\Phi(\beta,\bm{r})$,  where
$\beta=1/k_{\rm B} T$.
The bosons interact with each other by the repulsive contact interaction
$U>0$. $\mu$ is the  chemical potential.
In addition to the usual $\mathrm{U(1)}$ symmetry corresponding to
the conservation of the number of the bosons, the system is invariant
with respect to the global $\mathrm{SU(2)}$ transformation of the boson
field 
$\Phi \to \Phi'(\tau,\bm{r})={\cal U}\Phi(\tau,\bm{r})$,
where ${\cal U}$ is an $\mathrm{SU(2)}$ matrix.
As usual, the trap potential $V(\bm{r})$ has been included in the
definition of the action or the corresponding Hamiltonian.

Next, we introduce the dimensionless parameters and fields by rescaling:
\begin{align}
 & \Psi(\tilde{\tau},\tilde{\bm{r}})
   = \xi^{3/2}\Phi(\mu\tau,\xi^{-1}\bm{r}),
 \\
 & \tilde{V}(\tilde{\bm{r}})
   =\mu^{-1}V(\xi^{-1}\bm{r}),
 \\
 & \tilde{U}
   =\mu^{-1}\xi^{-3}U,
\end{align}
where $\xi=\hbar/\sqrt{m\mu}$ is the constant length scale. 
Effective action (\ref{action}) may be written now in terms of the
dimensionless parameters as
\begin{align}
 S
 &
 =\int_{0}^{\mu\beta}\!\!\!d\tilde{\tau}
  \!\!\int\!\!\!d^3\tilde{\bm{r}} \
  {\cal L}
 \nonumber \\
 {\cal L}
 &
 = \Psi^{\dagger}
    \left(
     \partial_{\tilde{\tau}}
     -\frac{1}{2}\tilde{\nabla}^2
     -1
     +\tilde{V}(\tilde{\bm{r}})
    \right)
   \Psi
   +\frac{\tilde{U}}{2}
    \left(
     \Psi^{\dagger}\Psi
    \right)^2.
\end{align}

Let us focus on classical field configurations independent of the
imaginary time. It will prove convenient to represent the
$\tau$-independent field $\Psi_c(\tilde{\bm{r}})$ by an amplitude
$f(\tilde{\bm{r}})$ and a two-component complex spinor configuration
$a(\tilde{\bm{r}})$:
$\Psi_c(\tilde{\bm{r}})=f(\tilde{\bm{r}})a(\tilde{\bm{r}})$. The spinor
$a(\tilde{\bm{r}})$ is normalized as
$a^{\dagger}(\tilde{\bm{r}})a(\tilde{\bm{r}})=1$. In terms of $f$ and
$a$, we can rewrite the effective action for the $\tau$-independent
field configuration $\Psi_c$, 
\begin{align}
 S_c
 &= \mu\beta
    \int\!\!\! d^3\tilde{\bm{r}}\
    {\cal L}_c
 \nonumber \\
 {\cal L}_c
 &=  \frac{1}{2}
     \left(
      \tilde{\nabla}
      f
     \right)^2
     +\frac{1}{2}
      f^2
       \tilde{\nabla}a^{\dagger}
      \cdot
       \tilde{\nabla}a
      +(\tilde{V}-1)f^2
      +\frac{\tilde{U}}{2}
       f^{4},
 \label{action2}
\end{align}
where we have used the relation
$\tilde{\nabla}(a^{\dagger}a)=(\tilde{\nabla}a^{\dagger}) a+a^{\dagger}(\tilde{\nabla}a)=\bm{0}$
deduced from the normalization condition for the spinor
$a(\tilde{\bm{r}})$. The stationary state also needs to satisfy the boundary condition
\begin{equation}
 \lim_{|\tilde{\bm{r}}|\to \infty}
 |\tilde{\bm{r}}|^2
 f(\tilde{\bm{r}})^2
 \tilde{\nabla}
 a(\tilde{\bm{r}})
 =0.
 \label{bc}
\end{equation}
It guarantees the stability of the solution with respect to small
rotations of the spinor $a(\tilde{\bm r})$ at the infinitely remote
boundary of the system.

Let us take the variation in action~(\ref{action2}) around the
classical  field $\Psi_c(\bm{r})$ using
the following expressions for the amplitude and the spinor configuration:
\begin{subequations}
\begin{align}
 & f(\tilde{\bm{r}})
   =f_0(\tilde{\bm{r}})
    +\delta f(\tilde{\bm{r}}),
 \\
 & a(\tilde{\bm{r}})
   =a_0(\tilde{\bm{r}})
    +\delta a(\tilde{\bm{r}}),
\end{align}
\label{variation}
\end{subequations}
where $\delta f(\tilde{\bm{r}})$ and $\delta a(\tilde{\bm{r}})$ are
the variations around $\Psi_c(\tilde{\bm{r}})$ for the density
profile and the spinor configuration, respectively. $f_0(\tilde{\bm{r}})$ and
$a_0(\tilde{\bm{r}})$ are the amplitude and the spinor of
some stationary field configuration.
In particular, $a_0 (\tilde{\bm{r}})$ will assume a form corresponding
to the skyrmion solution, which is to be defined shortly. Substituting
into the action, action (\ref{action2}) can be written as
\begin{equation}
 S_c
 =\mu\beta
  \int\!\! d^3\tilde{\bm{r}}
  \left[
   {\cal L}_0
   +{\cal L}_1
   +{\cal L}_2
   +\cdots
  \right],
\end{equation}
where ${\cal L}_{i}$ is the Lagrangian density related to the $i$th
order of the variation $\delta f(\tilde{\bm r})$ and
$\delta a(\tilde{\bm r})$. The Lagrangian densities
up to the second-order variation are then expressed as
\begin{widetext}
\begin{subequations}
\begin{align}
 {\cal L}_0
 &=\frac{1}{2}
   \left(
    \tilde{\nabla}
    f_0
   \right)^2
   +\frac{1}{2}
    f_0^2
    \tilde{\nabla}
    a_0^{\dagger}
    \cdot
    \tilde{\nabla}
    a_0
   +(\tilde{V}-1)
    f_0^2
   +\frac{\tilde{U}}{2}
    f_0^{4},
 \\
 {\cal L}_{1}
 &=\left[
    -\tilde{\nabla}^2
     f_0
    +
     \left\{
      \tilde{\nabla}a_0^{\dagger}
      \cdot
      \tilde{\nabla}a_0
      +2 \tilde{V}
      -2
     \right\}
     f_0
     +2\tilde{U}f_0^3
   \right]
   \delta f
   -\frac{1}{2}
   \left[
    \tilde{\nabla}
    \cdot
    \left(
     f_0^2
     \tilde{\nabla}
     a_0^{\dagger}
    \right)
    \delta a
    +h.c.
   \right],
 \label{first_variation}\\
 {\cal L}_{2}
 &=\delta f
   \left[
    -\frac{1}{2}
     \tilde{\nabla}^2
    +
    \tilde{V}
    -1
    +3\tilde{U}
     f_0^2
    -\frac{3}{2}
     \tilde{\nabla}
     a_0^{\dagger}
     \cdot
     \tilde{\nabla}
     a_0
   \right]
   \delta f
   +\frac{1}{2}
   \left[
    f_0
    \left(
     \tilde{\nabla}
     \delta a^{\dagger}
    \right)
    +2
     \left(
      \tilde{\nabla}
      a_0^{\dagger}
     \right)
     \delta f
   \right]
   \cdot
   \left[
    f_0
    \left(
     \tilde{\nabla}
     \delta a
    \right)
    +2
     \left(
      \tilde{\nabla}
      a_0
     \right)
     \delta f
   \right].
 \label{second_variation}
\end{align}
\end{subequations}
\end{widetext}

\subsection{Mapping to a problem in classical mechanics}

Setting $\int\!\! d^3\tilde{\bm r}{\cal L}_1=0$ for any
$\delta f(\tilde{\bm r})$ and $\delta a(\tilde{\bm r})$ leads to the
Euler-Lagrange equations giving the extremum of the action:
\begin{align}
 & -\tilde{\nabla}^2
    f_0
   +2
    \left[
     \frac{1}{2}
     \tilde{\nabla}a_0^{\dagger}
      \cdot
     \tilde{\nabla}a_0
     +\tilde{V}
     -1
    \right]
    f_0
   +2\tilde{U}f_0^3
   =0
 \label{ELeq1} \\
 & 2\tilde{\nabla}
    \cdot
    \left(
     f_0^2
     \tilde{\nabla}a_0
    \right)
 \notag \\
 & =\left[
     a_0^{\dagger}
     \tilde{\nabla}
     \cdot
     \left(
      f_0^2
      \tilde{\nabla}a_0
     \right)
     +\left\{
      \tilde{\nabla}
      \cdot
      \left(
       f_0^2\tilde{\nabla}a_0^{\dagger}
      \right)
      \right\}
      a_0
    \right]a_0.
 \label{ELeq2}
\end{align}
The latter equation follows by recalling that $a^\dagger a=1$, so that
$a_0^{\dagger}\delta a + \delta a^\dagger a_0=0$.

For simplicity, we further assume that the external
potential is spherically symmetric,
$\tilde{V}(\tilde{\bm{r}})=\tilde{V}(\tilde{r})$, where
$\tilde{r}=|\tilde{\bm{r}}|$. Then, in terms of the spinor
configuration, we can adopt the most general ansatz with the same
spherical symmetry~\cite{MS}, 
\begin{align}
 f_0(\tilde{\bm{r}})
 &=f_0(\tilde{r}),
 \label{ansatz1} \\
 a_0 (\tilde{\bm{r}})
 &=
    \begin{pmatrix}
     \sin\omega(\tilde{r})\cos\theta
     -i\cos\omega(\tilde{r})
     \\
     \sin\omega(\tilde{r})
     \sin\theta
     e^{-i\phi}
    \end{pmatrix}
    ,
 \label{ansatz2}
\end{align}
where
$\tilde{\bm{r}}=(\tilde{r}\sin\theta\cos\phi,\tilde{r}\sin\theta\sin\phi,\tilde{r}\cos\theta)$.
At the infinitely remote boundary, we impose
$\omega(\tilde{r}=\infty)=N\pi$, where $N$ is any integer.
If we additionally adopt the boundary condition $\omega(\tilde{r}=0)=0$,
and $\omega(\tilde{r})$ changes from $0$ to $N\pi$ as $\tilde{r}$
changes from $0$ to infinity, ansatz (\ref{ansatz2}) means that the
spinor configurations wraps the three-dimensional sphere $S^3$ $N$
times, and thus corresponds to the skyrmion solution. Hereafter, we
restrict the discussion to the simplest case of $N=1$.

Let us then consider Euler-Lagrange equations (\ref{ELeq1}) and
(\ref{ELeq2}) for $N=1$ skyrmion (\ref{ansatz1}) and (\ref{ansatz2}). 
Substituting Eqs.~(\ref{ansatz1}) and (\ref{ansatz2}) into
Euler-Lagrange equations (\ref{ELeq1}) and (\ref{ELeq2}), we obtain the
following differential equations:
\begin{align}
 & -\frac{d^2f_0}{d\tilde{r}^2}
   -\frac{2}{\tilde{r}}
    \frac{df_0}{d\tilde{r}}
 \nonumber \\
 & \quad
   +2
    \left[
     \frac{1}{2}
     \left(
      \frac{d\omega}{d\tilde{r}}
     \right)^2
     +\left(
       \frac{\sin\omega}{\tilde{r}}
      \right)^2
     +\tilde{V}
     -1
    \right]
    f_0
   +2\tilde{U}f_0^3
   =0.
 \label{ELeq1v2} \\
 & \frac{d^2\omega}{d\tilde{r}^2}
   +\left[
     \frac{2}{\tilde{r}}
     +\frac{2}{f_0}
      \frac{df_0}{d\tilde{r}}
    \right]
    \frac{d\omega}{d\tilde{r}}
   -\frac{\sin{2\omega}}{\tilde{r}^2}
   =0
 \label{ELeq2v2}
\end{align}
To analyze these  differential equations, it is very convenient to
change the variable from $\tilde{r}$ to $t=\ln\tilde{r}$. Then, as
$0<\tilde{r}<\infty$ , $-\infty<t<\infty$. Rewriting Eq.~(\ref{ELeq2v2})
in terms of $t$, we obtain 
\begin{equation}
 \frac{d^2\bar{\omega}}{dt^2}
 =-\frac{dW}{d\bar{\omega}}
  -\eta(t)
   \frac{d\bar{\omega}}{dt},
\label{classical_EOM}
\end{equation}
where $\bar{\omega}(t)=\omega(e^t)=\omega(\tilde{r})$.
This equation
can be regarded as describing classical motion of a particle in
an external potential with dissipation. $W(\bar{\omega})$ and $\eta(t)$,
respectively,
correspond to the potential energy and the ``time''-dependent
dissipation, given by
\begin{align}
 & W(\bar{\omega})
   =\frac{1}{2}\cos{2\bar{\omega}},
 \label{W} \\
 & \eta(t)
   =1+\frac{d}{dt}\ln\bar{f}_0^2,
 \label{eta}
\end{align}
where $\bar{f}_0(t)=f_0(e^t)=f_0(\tilde{r})$. In terms of $N=1$ skyrmion
solution, boundary conditions for $\omega(\tilde{r})$ may be written as
$\omega(0)=d\omega/d\tilde{r}(0)=0$ at the origin of the
three-dimensional space, and $\omega(\infty)=\pi$,
$d\omega/d\tilde{r}=0$ at the infinitely remote boundary. These boundary
conditions in terms of $t$ translate into 
\begin{equation}
 \left\{
  \begin{aligned}
   & \bar{\omega}(-\infty)=0,
   & \frac{d\bar{\omega}}{dt}(-\infty)=0,
   \\
   & \bar{\omega}(\infty)=\pi,
   & \frac{d\bar{\omega}}{dt}(\infty)=0.
  \end{aligned}
 \right.
 \label{bcs}
\end{equation}

Integrating equation of motion (\ref{classical_EOM}) with respect to
$t$, and imposing the set of boundary conditions (\ref{bcs}), one
obtains the following necessary condition for the existence of the
skyrmion solution: 
\begin{equation}
 \int_{-\infty}^{\infty}\!\!\!dt\
 \eta(t)
 \left(
  \frac{d\bar{\omega}}{dt}
 \right)^2
 =0.
 \label{integrated_condition}
\end{equation}
The condition implies that the total integrated dissipation in the
problem vanishes \cite{Landau}. It is now clear that a skyrmion solution
exists only when the density profile takes a special form, so that the
solution  $\bar{\omega}(t)$ satisfies
condition~(\ref{integrated_condition}).

\subsection{Stability of the skyrmion and a quantum-mechanical
eigenvalue problem}

The integral condition provides only the necessary condition for the
existence of the skyrmion solution, but does not guarantee its
stability. Here, we analyze the second-order variation, to obtain a
further  condition for the stable skyrmion solutions.

Let us assume that an appropriate external potential
$\tilde{V}(\tilde{r})$ is given so that the Euler-Lagrange equations
allow a skyrmion solution $f_0(\tilde{r})$ and $a_0(\tilde{\bm{r}})$.
For the obtained skyrmion solution to be stable against local
variations, the second-order variation in
Lagrangian~(\ref{second_variation}) has to be positive. 
Obviously, the second term in Eq.~(\ref{second_variation})
is always positive. The positivity of the first term in
Eq.~(\ref{second_variation}) for arbitrary $\delta f(\tilde{\bm r})$
and $\delta a(\tilde{\bm{r}})$ is thus a sufficient condition for the
local stability of the skyrmion. 
The first term being a quadratic form of $\delta f$, its positivity for
an arbitrary $\delta f$ is equivalent to positive definiteness of the
linear operator 
\begin{equation}
 H =-\frac{1}{2}
   \tilde{\nabla}^2
 +V_{\rm eff}(\tilde{\bm r}),
 \label{second_variation_hamiltonian}
\end{equation}
where
\begin{equation}
 V_{\rm eff}(\tilde{\bm r})
  = \tilde{V}(\tilde{r})
    -1
    +3\tilde{U}f_0^2(\tilde{r})
    -\frac{3}{2}
     \tilde{\nabla}a_0^{\dagger}
      \cdot
     \tilde{\nabla}a_0 .
 \label{effective_potential}
\end{equation}
Positive definiteness of $H$ means that all the eigenvalues of $H$ are
positive. 
In fact, $H$ may be interpreted as the {\em quantum-mechanical}
Hamiltonian for a single particle in the external potential 
$V_{\rm eff}$. 
The (sufficient) condition for the stability
then corresponds to the ground-state energy of the
Hamiltonian being positive.

The external potential $\tilde{V}(\tilde{\bm r})$
depends on  both the spinor part and the density profile of
the skyrmion, as set by the Euler-Lagrange equations.
According to Eq.~(\ref{ELeq1}), $V_{\rm eff}(\tilde{\bm r})$
is also spherically symmetric.

We should mention that even if the ground-state energy of
Hamiltonian~(\ref{second_variation_hamiltonian}) is negative, it is
possible that second-order variation~(\ref{second_variation}) is still
positive, if the positive second term is large enough. 
We, however, will be unable to say more about this issue, and our
discussion will be limited to the sufficient condition for the skyrmion
stability formulated above.

\section{Construction of skyrmion solutions for trapped BECs}
\label{construction}

\subsection{Odd-function ansatz}
\label{odd-functionansatz}

Let us discuss a few concrete examples of skyrmion solutions.
In a usual formulation, we seek a solution for a given trap potential
$V(\bm{r})$. 
However, as discussed in Ref.~\cite{HO}, a generic trap potential does
not allow a skyrmion solution.
Thus, we solve the problem backward: first we determine the dissipation
$\eta(t)$ so that equation of motion~(\ref{classical_EOM}) has a
solution $\bar{\omega}(t)$ which represents a skyrmion. 
The assumed dissipation determines the density profile of bosons.
Finally, the trap potential $V(r)$ is determined so that it reproduces
the chosen $\eta(t)$. 

For convenience, here we introduce the new variable
\begin{equation}
y \equiv \baromega - \frac{\pi}{2}.
\end{equation}
Equation of motion~(\ref{classical_EOM}) is then rewritten as
\begin{equation}
 \frac{d^2 y }{dt^2}
 =   -  \eta(t) \frac{d y}{dt} - \sin{2y}.
 \label{EOM_for_y}
\end{equation}
Boundary conditions~(\ref{bcs}) read,
in terms of $y$,
\begin{align}
   & y(-\infty)=  - \frac{\pi}{2} , \mbox{\hspace*{1cm}}
   \frac{dy}{dt}(-\infty)=0,
\label{bc-y-minus}
   \\
   & y(\infty)= + \frac{\pi}{2} , \mbox{\hspace*{1cm}}
   \frac{dy}{dt}(\infty)=0.
\label{bc-y-plus}
\end{align}
We observe that the boundary conditions at $t=+\infty$ would be
automatically met if $y$ obeys boundary conditions~(\ref{bc-y-minus}) at
$t=-\infty$ {\em and} $y(t)$ is an odd function:
\begin{equation}
y(t) = - y(-t).
\label{oddy}
\end{equation}
This is sufficient to satisfy the original conditions~(\ref{bc-y-minus})
and~(\ref{bc-y-plus}), but not necessary.
However, here we focus on finding the solutions $y(t)$ which are odd in
$t$, because it is easier than solving the general problem.

Equation~(\ref{EOM_for_y}) is invariant under
\begin{equation}
 t  \to -t, \;\;\;\;\;\;\;
 y  \to -y,
\end{equation}
if $\eta(t)$ is also odd in $t$.
Thus we choose an odd $\eta(t)$ so that we can find an odd solution
$y(t)$. 
Although this is a somewhat restrictive choice, it is a useful ansatz in
construction of skyrmion solutions.

However, it turns out that some choice of odd $\eta(t)$
actually does not allow
an odd solution $y(t)$ which satisfies
the boundary conditions at $t = -\infty$.
Roughly speaking, if the dissipation is too strong, the particle
which starts at $y=-\pi/2$ with a vanishing speed at $t=-\infty$
can not reach $y=0$ at $t=0$.
We will examine two forms of odd $\eta(t)$ as examples:
\begin{align}
 \eta(t)=& n [ \theta(-t-1) - \theta(t-1) ] ,
\label{dissipation-step}
\\
 \eta(t)=& - n\tanh t,
\label{dissipation}
\end{align}
where $n$ is a positive parameter.
As we will demonstrate, for each case there is a critical parameter
$n_c$; an appropriate odd solution exists only if $n<n_c$.

The existence of an odd solution can be discussed in the following
manner. 
First we impose boundary conditions~(\ref{bc-y-minus}) only at
$t=-\infty$.  
Of course, they do not completely fix a solution but allow a family of
different solutions. 
This is evident by recalling that the trivial solution $y(t)=-\pi/2$
satisfies Eq.~(\ref{bc-y-minus}).

Then we attempt to find, among the solutions, the one which satisfies
$y(0)=0$. 
This would be the desired odd solution. 
The particle falls off the hill and approaches the potential minimum
$y=0$. 
If the particle is still rolling down the hill ($-\pi/2 < y < 0$) when
$t=0$, the particle is only accelerated by the negative friction for
$t>0$ and it must eventually reach $y=0$ at some positive $t$.
Namely, for any dissipation strength, there is a solution which
satisfies $y(t)=0$ at a positive $t$. 
On the other hand, while $y<0$, the velocity $dy/dt$ should always be
positive. Thus the smallest solution $t$ of $y(t)=0$ changes
continuously. 
Therefore, if there is another solution which satisfies $y(t)=0$ at a
negative $t$, there must be an odd solution with $y(0)=0$ thanks to the
intermediate value theorem. 

\subsection{Constant-$\eta$ regime}
\label{constatnetaregime}

In either case of Eq.~(\ref{dissipation-step}) or~(\ref{dissipation}),
we observe that, for $t \ll -1$,
\begin{equation}
 \eta(t) \sim n .
\end{equation}
This simply represents the constant-dissipation coefficient. 
The equation of motion
\begin{equation}
 \frac{d^2 y }{dt^2} =  - n \frac{d y}{dt}- \sin{2y}.
\label{EOM_for_y_c}
\end{equation}
in this regime is independent of time.
Thus, for any solution $y(t)$,
the translated solution $y(t+\tau)$
is also a solution for any $\tau$.
Let us discuss the solutions of this equation.
This will turn out to be useful in
determining the critical parameter $n_c$ for the
original equation of motion~(\ref{EOM_for_y}).

As we will see later, when $n \sim n_c$,
the particle comes very close to the minimum
($|y| \ll 1$) while still in the constant-$\eta$ regime ($t \ll -1$).
For small $y$, we may use the linearized equation of motion
\begin{equation}
 \frac{d^2 y }{dt^2} =   - n \frac{d y}{dt}- 2y,
\label{EOM_linear_y_c}
\end{equation}
instead of the full nonlinear equation~(\ref{EOM_for_y_c}).
Its solution can be easily obtained as
\begin{equation}
 y(t) = A_1 e^{-\lambda_1 (t+\tau)} + A_2 e^{-\lambda_2 (t+\tau)},
\label{linear_sol_y_c}
\end{equation}
where
\begin{align}
\lambda_1 =& \frac{n - \sqrt{n^2-8}}{2}, \\
\lambda_2 =& \frac{n + \sqrt{n^2-8}}{2} .
\end{align}
One can choose an arbitrary large positive $\tau$, thanks to the
translation invariance in time.
On the other hand, taking a large negative $\tau$
(for a fixed $t$) makes $y$
large, and may invalidate linear approximation~(\ref{EOM_linear_y_c}).

When $n < 2 \sqrt{2}$, $\lambda_{1,2}$ are a complex conjugate pair
and the solution represents a damped harmonic oscillation.
In this case, there is obviously a solution which reaches
$y=0$ within Eq.~(\ref{EOM_for_y_c}).
Namely, there is
a solution which satisfies $y(t)=0$ at a negative $t$.
As we discussed in Sec.~\ref{odd-functionansatz},
the intermediate value theorem assures that
there is an odd solution with $y(0)=0$ in this case.
This means that $n_c > 2 \sqrt{2}$.

Thus, in the following, we focus on $n > 2 \sqrt{2}$.
Then, the first term $\propto e^{-\lambda_1 (t+\tau)}$
is the leading one in the $\tau \to \infty$.
The second term $e^{-\lambda_2 (t +\tau)}$ vanishes
more quickly, but the subleading contribution
determines the critical exponent as we will show later.

In fact, in discussing the subleading contribution,
we must also consider the nonlinear effects
which were ignored in Eq.~(\ref{EOM_linear_y_c}).
We introduce the scaling of $y$ by the replacement
$y \to \alpha y$.
We are interested in the limit in which the original $y$
is small, namely $\alpha \to 0$.
The equation of motion now reads
\begin{equation}
 \frac{d^2 y }{dt^2}
 =   - n \frac{d y}{dt}- \frac{1}{\alpha} \sin{(2\alpha y)}.
\end{equation}
Considering the limit $\alpha \to 0$ and retaining
only the leading nonlinear term, we obtain
\begin{equation}
 \frac{d^2 y}{dt^2}
 =   - n \frac{d y}{dt}- 2 y + \frac{4}{3} \alpha^2 y^3 +\mathrm{O}(\alpha^3).
\end{equation}
We consider a series expansion of the solution $y$
in terms of $\alpha$, which can be regarded as a
perturbative expansion of nonlinear effects.

The lowest order $y^{(0)}$ is given by the solution
of linearized equation~(\ref{EOM_linear_y_c}).
The next order $y^{(1)}$ is of $O(\alpha^2)$, and is
given by a solution of
\begin{equation}
 \frac{d^2 y^{(1)}}{dt^2}
    + n \frac{d y^{(1)}}{dt} + 2 y^{(1)} =
\frac{4}{3} \alpha^2 (y^{(0)})^3  .
\end{equation}
This is an inhomogeneous linear differential equation
on $y^{(1)}$ for a given $y^{(0)}$,
which can be solved by a standard method.
Taking the $y^{(0)}$ as general solution~(\ref{linear_sol_y_c})
of the linear equation, we find the special solution
\begin{equation}
y^{(1)}(t) =
- \alpha^2
\frac{2 {A_1}^3}{3 \lambda_1 (\lambda_2 - 3 \lambda_1)} e^{-3\lambda_1(t+\tau)}
+\ldots ,
\label{y-nonlinear1}
\end{equation}
where only the leading term is given.
The general solution also contains solutions of
the corresponding homogeneous equation.
However, they have the same form as $y^{(0)}$ and can be
ignored in the following.

Combining with the solution of the linearized version
[Eq.~(\ref{linear_sol_y_c})], the leading and next-leading
terms in the $\tau \to \infty$ limit can be written as
\begin{equation}
 y(t) = A_1 e^{-\lambda_1 (t+\tau)} (1 + C e^{-\Delta \lambda (t+\tau)}),
\label{y_asymptotic}
\end{equation}
where
\begin{equation}
 \Delta \lambda \equiv \min{(\lambda_2 - \lambda_1, 2 \lambda_1)} .
\end{equation}
Namely, when $\lambda_2 > 3 \lambda_1$, the next-leading
term comes from the nonlinear effect instead of
the term proportional to $e^{-\lambda_2 (t+\tau)}$.

Recalling that the particle approaches $y=0$ from
$y=-\pi/2$, $A_1<0$.
The constant $C$ is determined by the solution of nonlinear equation of
motion~(\ref{EOM_for_y_c}) with the constant dissipation and boundary
conditions~(\ref{bc-y-minus}).
The solution gives the effective initial conditions
for the linearized equation.

\begin{figure}[tbp]
 \begin{center}
  \includegraphics[scale=0.65]{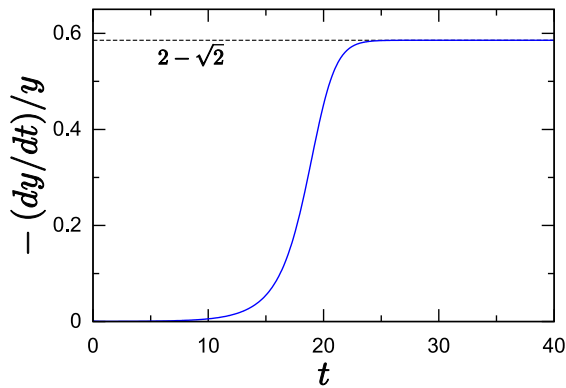}
  \caption{
    (Color online)
    The ratio $-(dy/dt)/y$ as a function of time $t$,
    within constant-$\eta$ equation~(\ref{EOM_for_y_c})
    and with boundary conditions~(\ref{bc-y-minus}).
    In the actual numerical calculation, a small initial
    velocity is given instead of zero in Eq.~(\ref{bc-y-minus})
    to avoid getting only the trivial solution.
    The result is robust against the change in the small
    initial velocity.
    In the figure, we show the result for $\eta=4$ as example.
    The ratio $-(dy/dt)/y$ increases monotonically, asymptotically
    approaching $\lambda_1 = 2 - \sqrt{2} \sim 0.585786 $.
    We obtained similar results for other $\eta > 2 \sqrt{2}$.
}
\label{dydt}
\end{center}
\end{figure}

Numerically solving Eq.~(\ref{EOM_for_y_c}) with Eq.~(\ref{bc-y-minus}),
we find that, the ratio $-(dy/dt)/y$
increases monotonically, as shown in Fig.~\ref{dydt},
when $n>2\sqrt{2}$.
This observation implies that $C<0$
in linear regime~(\ref{y_asymptotic}).

In fact, when the nonlinear effect gives the
next-leading term, it follows from Eq.~(\ref{y-nonlinear1}) that 
\begin{equation}
 C= - \alpha^2 \frac{2 {A_1}^2}{3 \lambda_1 (\lambda_2 - 3 \lambda_1)} <0.
\end{equation}
When the subleading contribution within the linear equation
gives the next-leading term, we do not have a proof
but $C<0$ seems certain from numerical results.

Furthermore, the numerical results indicate that $C$ is sufficiently
small so that $y$ never reaches $0$, when asymptotic
expression~(\ref{y_asymptotic}) is valid. 
This, of course, does not mean that there is no solution satisfying
$y(0)=0$ in the original Eq.~(\ref{EOM_for_y}), in which
the dissipation is turned off around $t \sim 0$.

\subsection{Critical parameters in the step-function case}
\label{criticalparameterinthestepfunctioncase}

Let us consider the step-function case
of Eq.~(\ref{dissipation-step}).
Here, the solution of Eq.~(\ref{EOM_for_y_c})
discussed in Sec.~\ref{constatnetaregime}
gives the ``initial condition'' at $t=-1$ for the equation
without the dissipation.
If the particle is very close to the minimum ($|y| \ll 1$) at
$t=-1$, the successive motion is just a harmonic oscillation
with the angular frequency of $\sqrt{2}$.
We will show later that for $n \sim n_c$,
$|y| \ll 1$ indeed holds at $t=-1$.

The phase $\zeta$ of the oscillation is given as
\begin{equation}
 \tan{\zeta} = y \sqrt{2} \left(\frac{d y}{dt}\right)^{-1} .
\end{equation}
When $n \to n_c$, the particle just manages to reach $y=0$ at $t=0$.
For that, we need to give the optimal initial condition at $t=-1$,
namely, $-(dy/dt)/y$ with the maximum possible value.
As discussed in Sec.~\ref{constatnetaregime}, $-(dy/dt)/y$ for
Eq.~(\ref{EOM_for_y_c}) monotonically increases.
Thus the optimal initial condition [maximum possible $-(dy/dt)/y$]
is realized by letting the particle
spend infinite time around $y\sim 0$ before $t=-1$,
namely, by taking $\tau \to \infty$.
In this limit,
the first term in Eq.~(\ref{y_asymptotic}) dominates
and the initial condition at $t=-1$ is given by
\begin{equation}
 \tan{\zeta_0} = - \frac{\sqrt{2}}{\lambda_1}.
\label{zeta0_optimal}
\end{equation}
We emphasize that $T$ depends only on the ratio
between the ``initial'' velocity and ``initial'' coordinate on $t=-1$,
which converges to a finite value in the limit $\tau \to \infty$.
The time required to reach the minimum ($y=0$) in the
harmonic oscillation is
\begin{equation}
 T = \frac{|\zeta_0|}{\sqrt{2}} =
\frac{1}{\sqrt{2}} \tan^{-1} \frac{\sqrt{2}}{\lambda_1} .
\end{equation}

The critical dissipation coefficient $n_c$ in this problem is
thus given by
\begin{equation}
 T = \frac{1}{\sqrt{2}} \tan^{-1} \frac{\sqrt{2}}{\lambda_1} = 1,
\end{equation}
so that the particle arrives at $y=0$ at $t=0$.
Therefore we find
\begin{equation}
  n_c = \frac{2 \sqrt{2}}{\sin{2 \sqrt{2}}} \sim 9.18107 \ldots
\label{n_c-step}
\end{equation}
In the limit $n \to n_c$, $\tau \to \infty$, and
thus both $y(-1)$ and $dy/dt(-1)$ vanish.
This implies that the velocity of the particle when it reaches
the potential minimum, $dy/dt(0)$, also vanishes.
Let us discuss its critical behavior, namely, how
$dy/dt(0)$ depends on $n_c - n$ when $n \lesssim n_c$.

For $n<n_c$, if we take $\tau \to \infty$
the particle reaches the minimum before $t=0$ because $T<1$.
By letting the particle spend less time around $y \sim 0$ by
taking a smaller $\tau$,
we can change the initial condition at $t=-1$ so that $\zeta_0$
is smaller than the optimal value~(\ref{zeta0_optimal}).
Therefore, for $n<n_c$,
there is a solution which reaches the minimum
at $t=0$.
Combining
\begin{equation}
  - \frac{1}{\sqrt{2} y}\left(\frac{dy}{dt}\right)(t=-1)
= \tan{\frac{1}{\sqrt{2}}}
\end{equation}
with Eq.~(\ref{y_asymptotic}), for small $n_c -n$ we find
\begin{equation}
 \left. \frac{\partial \lambda_1}{\partial n}\right|_{n=n_c} (n-n_c)
+ (\lambda_2-\lambda_1) C e^{- \Delta \lambda (\tau-1)} \sim 0.
\end{equation}
This implies that
\begin{equation}
e^{- \Delta \lambda \tau} \propto n_c -n.
\end{equation}
As a consequence,
\begin{equation}
 y(-1) \propto  \frac{dy}{dt}(-1) \propto e^{- \lambda_1 \tau}
\propto (n_c - n)^{\lambda_1 / \Delta \lambda},
\end{equation}
which leads to
\begin{equation}
 \frac{dy}{dt}(0) \propto (n_c - n)^{\lambda_1 / \Delta \lambda}.
\label{vcrit_general}
\end{equation}
In the present case, for $n\sim n_c$,
$\lambda_2 > 3 \lambda_1$, namely,
$\Delta \lambda = 2 \lambda_1$.
Thus we obtain
\begin{equation}
 \frac{dy}{dt}(0) \propto (n_c - n)^{1/2} .
\label{vcrit_step}
\end{equation}

For $n>n_c$, even under the optimal condition
$\tau \to \infty$, the particle cannot
reach $y=0$ at $t=0$.
In this case, the odd solution does not exist.

\subsection{Critical parameters in the tanh case}

Now let us consider $\tanh$ case~(\ref{dissipation}).
The mathematics is somewhat more complicated but
the physics is quite similar to the previous one.

As in the previous problem, for $n \sim n_c$, we can assume that
the particle comes very close to the minimum $y \sim 0$
at negative time.
The linearized equation of motion, with the full time dependence
of the dissipation, reads
\begin{equation}
 \frac{d^2 y }{dt^2}
 =   + n  \frac{d y}{dt}  \tanh{t} - 2y.
 \label{EOM_linear_y}
\end{equation}

This equation has the general solution
\begin{align}
y =& C_1 P_{n/2}^{(\sqrt{n^2-8}/2)}(\tanh{t})(1-\tanh^2{t})^{-n/4}
\notag \\
&+ C_2 Q_{n/2}^{(\sqrt{n^2-8}/2)}(\tanh{t})(1-\tanh^2{t})^{-n/4} ,
\label{y_in_PQ}
\end{align}
where $C_{1,2}$ are constants, and
$P_{\mu}^{(\nu)}$ and $Q_{\mu}^{(\nu)}$ are
associated Legendre functions.

Let us first discuss the asymptotic behavior of the above solution
in the limit $t \to -\infty$.
In this limit, equation of motion~(\ref{EOM_linear_y}) reduces
to Eq.~(\ref{EOM_linear_y_c}), and thus the solution should
be equivalent to Eq.~(\ref{linear_sol_y_c}).
In fact, using
\begin{equation}
 \tanh{t} \sim -1 + 2 e^{2t} + O(e^{4t}),
\end{equation}
and the asymptotic expansion of the associated Legendre
functions, we have confirmed that Eq.~(\ref{y_in_PQ})
coincides with Eq.~(\ref{linear_sol_y_c})
in the limit $t \to - \infty$.

Now, the similar discussion as in
Sec.~\ref{criticalparameterinthestepfunctioncase} applies here. 
Namely, the numerical solution implies that
$-(dy/dt)/y$ increases monotonically.
Thus, the optimal condition for reaching $y=0$
is realized when the particle spends infinite time around $y \sim 0$
before $t \sim 0$.
This can be done by replacing $t$ with $t + \tau$ and taking
$\tau \to \infty$.
In this limit, the asymptotic behavior of the
solution in the $t \ll -1$ regime is dominated by
$e^{-\lambda_1 (t + \tau)}$.

In the following, we demonstrate that
\begin{equation}
 n_c = 3 .
\end{equation}
To show that, let us set $n=3$.
The two independent solutions reduce to
\begin{align}
P_{3/2}^{(1/2)}(\tanh{t})(1-\tanh^2{t})^{-3/4}
 & \sim \frac{1}{2 \sqrt{2\pi}}e^{-\lambda_2 t}, \\
Q_{3/2}^{(1/2)}(\tanh{t})(1-\tanh^2{t})^{-3/4}
 & \sim \sqrt{\frac{\pi}{2}} e^{-\lambda_1 t},
\end{align}
in the limit $t \to -\infty$.
This implies that, under the optimal condition,
the solution consists only of the $Q_{3/2}^{(1/2)}$ term.

On the other hand,
for $n=3$, the Taylor expansion around $t=0$ reads
\begin{align}
P_{3/2}^{(1/2)}(\tanh{t})(1-\tanh^2{t})^{-3/4}
\sim& - \sqrt{\frac{2}{\pi}} + O(t^2),
\\
Q_{3/2}^{(1/2)}(\tanh{t})(1-\tanh^2{t})^{-3/4}
\sim&  - \sqrt{2 \pi} t + O(t^2) .
\end{align}
Namely, the solution consisting only
of the $Q$ term just crosses $y=0$ on $t=0$.
By perturbing the solutions around $n=3$,
it can be shown that the solution with $y(0)=0$
exists for $n<3$.
This means that $n=3$ is indeed the critical value $n_c$.

As in the case in Sec.~\ref{criticalparameterinthestepfunctioncase},
the velocity at the potential minimum
$dy/dt(0)$ vanishes as $n$ approaches $n_c$ from below.
The critical behavior can be obtained by a similar
argument, and is given by Eq.~(\ref{vcrit_general}).
In the present case, $\lambda_1=1$
and $\lambda_2 = 2 < 3 \lambda_1$.
Thus we find the linear behavior
\begin{equation}
 \frac{dy}{dt}(0) \propto (n_c - n),
\label{vcrit_tanh}
\end{equation}
when $n \lesssim n_c$.

\subsection{Requirement of a finite number of bosons}

We consider the situation where all the bosons are confined in
the finite space by external trap potential $V(\bm{r})$.
If it is to be realized in experiments,
the total number of the bosons should be finite.
This gives an additional requirement
independent of the stability.

In terms of the density profile ${f_0}^2$,
the condition of the finite number of bosons
is easily given as
\begin{equation}
 \int\!\!\!d\tilde{\bm{r}}\
 {f_0}^2({\tilde{r}})
 =4\pi
  \int_0^{\infty}\!\!\!d\tilde{r}\
  \tilde{r}^2 f_0^2(\tilde{r})<\infty.
 \label{trap_condition}
\end{equation}

For the choice of the $\tanh$ dissipation [Eq.~(\ref{dissipation})],
the density profile can be obtained via Eq.~(\ref{eta}) as
\begin{equation}
 f_0^2(\tilde{r})
 =B\frac{\tilde{r}^{n-1}}{\left(1+\tilde{r}^2\right)^n},
 \label{density_profile}
\end{equation}
where $B$ is a positive constant which appears from the integration of
Eq.~(\ref{eta}).
The density profile in the $n=1$ case was discussed
in Ref.~\cite{HO}.
$\tilde{r}^2f_0^2(\tilde{r})$ in the
integral of Eq.~(\ref{trap_condition})
asymptotically behaves as $\sim \tilde{r}^{-n+1}$ for large $\tilde{r}$.
Accordingly, for $n\le 2$ the total number of the trapped bosons
diverges, violating condition~(\ref{trap_condition}).
Hereafter, we focus on the finite-bosons case,
$n>2$.

\section{Numerical solution for a skyrmion}
\label{numerical}

\subsection{Solution of the equation of motion}

Equation of motion (\ref{classical_EOM})
determines $\bar{\omega}(t)$ for a given choice of $\eta(t)$.
While we have determined critical parameters in Sec.\ref{construction},
unfortunately, the full solution
cannot be obtained analytically.
Thus, here we solve Eq.~(\ref{classical_EOM}) numerically.
The numerical solution can be also used to check the
analytical predictions on the critical parameters
discussed in Sec.~\ref{construction}.

To reiterate, we seek a solution which satisfies the
boundary conditions at $t=\pm \infty$ [Eq.~(\ref{bcs})].
For an odd $\eta(t)$, which is the case we discuss in
this paper, such a solution satisfies Eq.~(\ref{oddy}).
To find the odd solution for an odd $\eta(t)$,
it is enough to require
\begin{equation}
 \bar{\omega}(0) = 0 ,
\end{equation}
together with either of the boundary conditions
at $t=-\infty$ or $t=\infty$ in Eq.~(\ref{bcs}).

\begin{figure}[tbp]
 \begin{center}
  \includegraphics[scale=0.65]{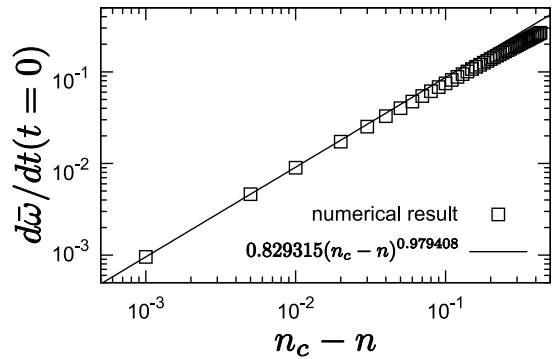}
  \caption{
  The velocity of the particle at $t=0$ as a function of $n_c - n$,
  for the tanh dissipation [Eq.~(\ref{dissipation})].
  The squares are numerical results obtained by the
  shooting method.
  The line is the best fit assuming a power law.
  The critical parameter $n_c$ is determined by the
  analytic prediction.
  The good fit to the power law means that the analytic
  prediction of $n_c$ is consistent with
  the numerical calculation.
  Moreover, the exponent $0.979\cdots$ obtained by the fit
  is also consistent with the analytic prediction of unity.
}
\label{v_tanh}
\end{center}
\end{figure}

\begin{figure}[tbp]
 \begin{center}
  \includegraphics[scale=0.65]{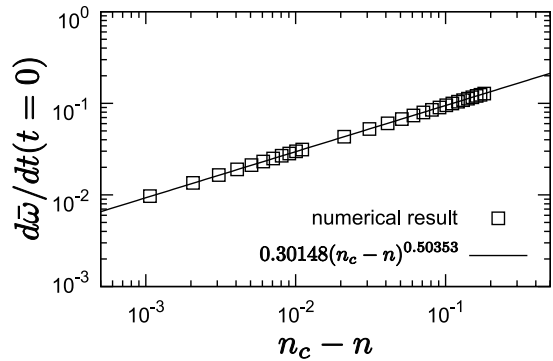}
  \caption{
  The velocity of the particle at the potential minimum ($t=0$)
  as a function of $n_c - n$,
  for the step function dissipation.
  The line is the best fit assuming a power law to
  the numerical results, which are shown as squares.
  The critical parameter $n_c$ is determined by the
  analytic prediction.
  The numerical result is again in good agreement
  with the predictions.
}
\label{v_step}
\end{center}
\end{figure}

Based on this observation, we adopt the so-called ``shooting method''
in the numerical scheme, which is explained in
Appendix~\ref{shootingmethod}.  
In Fig.~\ref{omega}, we show the numerical result for
the case of Eq.~(\ref{dissipation}).
\begin{figure}[tbp]
 \begin{center}
  \includegraphics[scale=0.65]{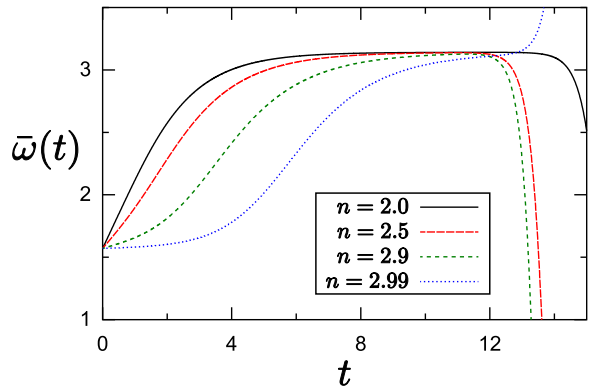}
  \caption{
  (Color online)
  Trajectories of $\bar{\omega}(t)$ in the case of
  $\eta(t)=-n\tanh{t}$. The trajectories in
  $t<0$ are antisymmetric to that in $t>0$, so we show
  $\bar{\omega}(t)$ only for $t>0$.
  Here, calculating the trajectories
  numerically, $\bar{\omega}(t=0)=\pi/2$ is imposed as an initial
  condition.
  Now, although we define $n>2$,
  the trajectory for $n=2$ is also shown.}
  \label{omega}
 \end{center}
\end{figure}
Here, we observe that
the velocity at $t=0$ vanishes
as the dissipation parameter $n$ approaches $n_c=3$.
This is indeed consistent with the analytic prediction
on the critical parameter $n_c=3$ and on the critical
behavior.
To see this more clearly, in Fig.~\ref{v_tanh},
we show the numerical result on $d\bar{\omega}/dt(0)
=dy/dt(0)$ as a function of $n_c -n$.
The result is in good agreement with
the analytic predictions.
We also have made a similar comparison for
step-function dissipation~(\ref{dissipation-step}) in
Fig.~\ref{v_step}, and found agreement
with analytic predictions~(\ref{n_c-step}) and~(\ref{vcrit_step}) as
well. 

\subsection{Stability of the skyrmion and minimal number of bosons}

We have shown that the skyrmion as a solution of
the Euler-Lagrange equation exists for $n<n_c$,
and obtained the solution numerically.
The spinor configuration $\bar{\omega}(t)$ is
given by the solution of a classical mechanics problem.
The density profile $f_0(\tilde{r})$ can then be obtained.

\begin{figure}[tbp]
 \begin{center}
  \includegraphics[scale=0.7]{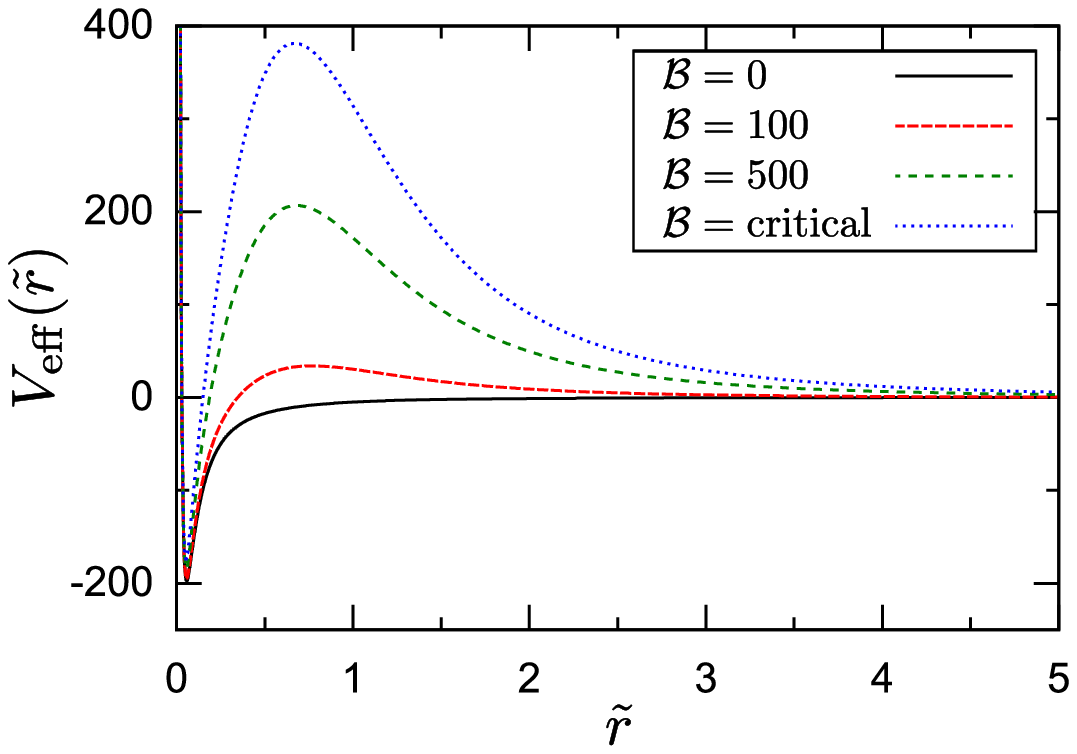}
  \includegraphics[scale=0.71]{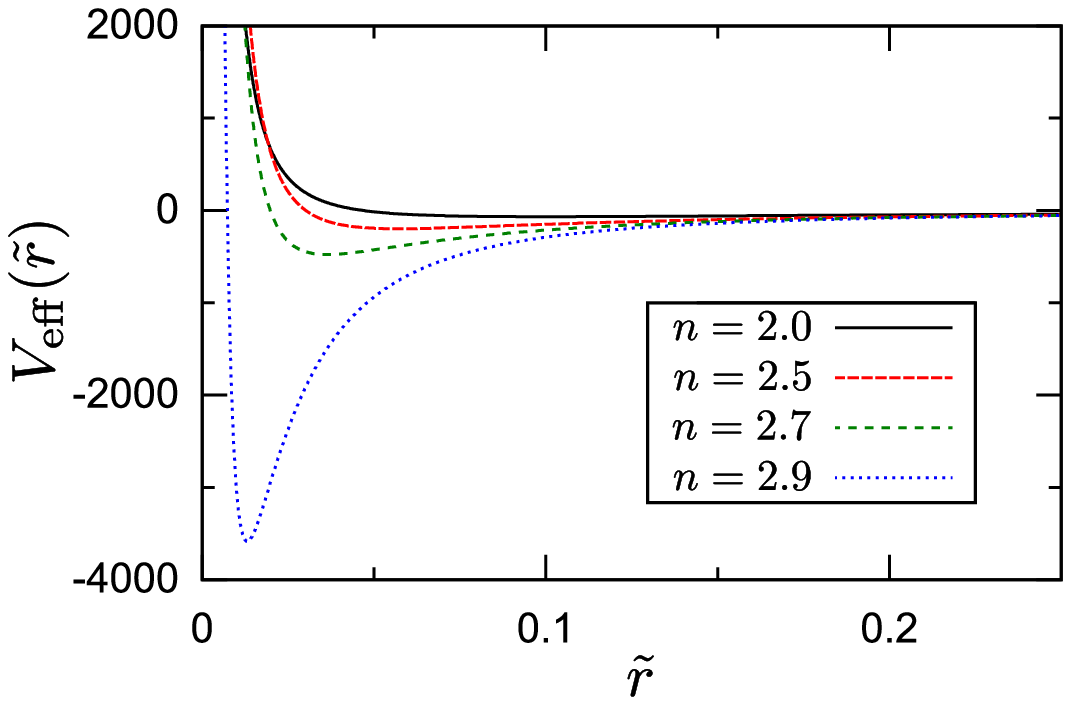}
  \caption{
  (Color online)
  Plots of the effective potential $V_{\rm eff}$
  defined in Eq.~(\ref{effective_potential}) for given
  $\eta(t)=-n\tanh{t}$.  In the upper graph,
  $V_{\rm eff}(\tilde{r})$ with fixed $n=2.5$ are plotted. In the lower graph,
  the $n$ dependence of $V_{\rm eff}(\tilde{r})$ with fixed ${\cal B}=0$ is
  shown. For each value of $n$, there is a critical
  value of ${\cal B}$ which makes the ground-state energy of
  Hamiltonian~(\ref{second_variation_hamiltonian})  exactly zero. In the
  region of ${\cal B}$ shown in the upper graph, the ground-state eigenvalues
  are negative. }
  \label{fig_effective_potential}
 \end{center}
\end{figure}

\begin{figure}[tbp]
 \begin{center}
  \includegraphics[scale=0.68]{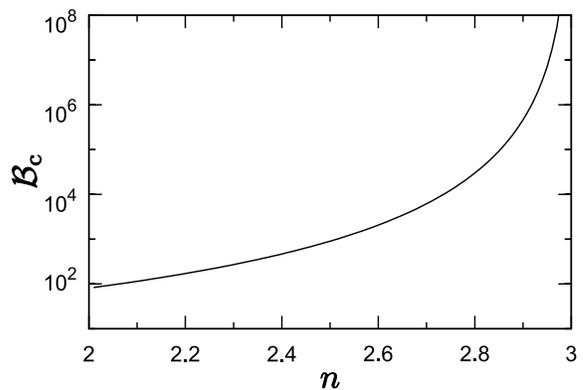}
  \caption{
  The critical value ${\cal B}_{\rm c}$ as a function of $n$. The
  values of ${\cal B}_{\rm c}$ are plotted on logarithmic scale. We
  find that ${\cal B}_{\rm c}$ increases with $n$. Eventually, in the
  limit $n\to3$, ${\cal B}_{\rm c}$ diverges.}
  \label{criticalB}
 \end{center}
\end{figure}

The next step is to examine the stability of the obtained skyrmion.
Note that there is a free parameter $B$ appearing in density
profile~(\ref{density_profile}). Substituting
the obtained $\bar{\omega}(t)$ and $\bar{f}_0(t)$ into Euler-Lagrange
equation~(\ref{ELeq2v2}), the trap potential $\tilde{V}(\tilde{r})$
realizing the skyrmion is obtained. Then, we find that 
$\tilde{V}(\tilde{r})$ depends on $\tilde{U}$ and $B$ as a function of
$\tilde{U}B={\cal B}$. This information on
$\omega(\tilde{r})$, $f_0(\tilde{r})$, and $\tilde{V}(\tilde{r})$
determines the effective
potential~(\ref{effective_potential}), shown in
Fig.~\ref{fig_effective_potential}.
In order to obtain the stable skyrmion, ${\cal B}$ larger than the
critical value ${\cal B}_{\rm c}$ is needed. Indeed, calculating the
ground-state energy of quantum-mechanical
Hamiltonian~(\ref{second_variation_hamiltonian}) by use of the numerical
diagonalization method, it is found that as $n$ increases, the critical
value ${\cal B}_{\rm c}$ diverges as $n$ approaches $n_c=3$,
as shown in Fig.\ref{criticalB}.

\begin{figure}[tbp]
 \begin{center}
  \includegraphics[scale=0.7]{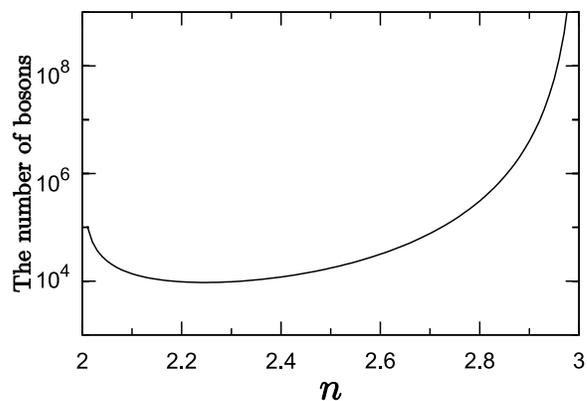}
  \caption{
  The $n$ dependence of the number of the trapped bosons in
  $B=\tilde{U}^{-1}{\cal B}_{\rm c}$ for a given $n$. It has the minimal
  value of $\sim 10^4$ around $n=2.25$.}
  \label{total_bosons}
 \end{center}
\end{figure}

Finally, let us investigate the total number of the trapped bosons for some
values of ${\cal B}$ and $n$. Due to Eq.~(\ref{density_profile}) for the
density profile, the number of
bosons monotonically increases as a function of ${\cal B}$: it is proportional
to ${\cal B}$. For a given $n$, the number of the trapped bosons is
minimal at ${\cal B}={\cal B}_{\rm c}$. The $n$ dependence of the
number of the bosons in ${\cal B}={\cal B}_{\rm c}$ is then shown in
Fig.~\ref{total_bosons}. We find the minimum at $n \sim 2.25$, with the
value of $\sim 10^4$. It is
comparable to the typical numbers in the experiments.

\section{Summary}
\label{summary}

In this paper, we have discussed the skyrmion
configurations of the two-component
spinor BECs confined by a trap potential.
The necessary condition for the existence of the skyrmion has
been formulated earlier by two of the authors~\cite{HO}.
The Euler-Lagrange equation, which must be satisfied by
the skyrmion, turned out to give an equation of motion
for a fictitious classical particle subject to time-dependent
dissipation.
A skyrmion solution satisfying appropriate boundary conditions
exists only for specially chosen trap potentials.
Some of those trap potentials were constructed by considering
dissipation which is an odd function of time.
It was expected to allow a solution of the equation of motion,
which is odd in time.

In this paper, we have developed this approach further.  
We have mainly considered two classes of solutions,
obtained by taking the dissipation as proportional to
the step function of time [Eq.~(\ref{dissipation-step})],
and to hyperbolic tangent of time [Eq.~(\ref{dissipation})].
We have found that, even though
these choices of dissipation are odd in time, they do not
allow a skyrmion solution if the dissipation is too strong.
We have obtained the critical parameter exactly, and
also determined a critical behavior in the velocity at
the potential minimum.
These predictions are verified by numerical solution.

Furthermore, we discussed the stability of the
skyrmion, taking into account the second-order variational theory.
In our formulation
the problem of the stability of the
skyrmion is mapped onto that of the sign of the lowest eigenvalue of
a certain quantum-mechanical Hamiltonian
determined by the skyrmion solution.

For the $\tanh$ model, the density profile and the trap potential
reproducing the skyrmion have several parameters $\tilde{U}$, $n$ and
$B$, and we determined the region in the parameter space that leads to a
stable skyrmion. For a given $n$, ${\cal B}=\tilde{U} B$
needs to be larger than a certain value for the skyrmion to be stable
(see Fig.\ref{criticalB}).

\begin{figure}[tbp]
 \begin{center}
  \includegraphics[scale=0.7]{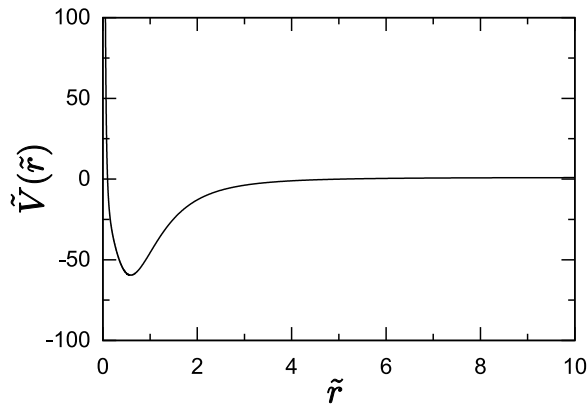}
  \caption{
  The example of the trap potential $\tilde{V}(\tilde{r})$
  reproducing the skyrmion. The density profile under this potential is
  given by Eq.(\ref{density_profile}) with $n=2.25$ and
  ${\cal B}=213.495$. As seen in Fig.~\ref{total_bosons}, for these
  $n$ and ${\cal B}$, the number of the trapped bosons is
  minimal ($\sim 10^4$).}
 \label{trap_potential}
 \end{center}
\end{figure}

We have found that there exists a minimal number of bosons
of $\sim 10^4$ allowing a stable skyrmion solution.
For illustration,
the trap potential required to reproduce the skyrmion with this minimal
number of particles is shown in Fig.~\ref{trap_potential}.
In order to connect our results to real experiments,
it would be important to see the asymptotic form of the trap
potential stabilizing the skyrmion. It can be analytically derived from
asymptotic analysis of differential equations~(\ref{ELeq1v2})
and~(\ref{ELeq2v2}). In the generic case that the odd $\eta(t)$
asymptotically reaching $\mp n$, respectively, as $t\to \pm\infty$, as
seen in Appendix~\ref{appendix:asymptotics}, the form of the trap
potential can be obtained as Eqs.~(\ref{asymptotic_potential1})
and~(\ref{asymptotic_potential2}), respectively, for $\tilde{r}\ll 1$
and $\tilde{r}\gg 1$.
One can easily find that these equations are consistent with the result
of the analytically solvable cases discussed in Ref.~\cite{HO}: 
$\eta(t)=0$ for $n=0$ and $\eta(t)=-\tanh{t}$ for $n=1$.
For $n>2$, the leading terms in Eqs.~(\ref{asymptotic_potential1}) and
(\ref{asymptotic_potential2}) are both proportional to 
$\sim 1/\tilde{r}^2$. Interestingly, the power of the leading terms 
is independent of $n$.
In particular, in the $n=2.25$ case shown in Fig.~\ref{trap_potential},
the analytically obtained asymptotic form for $\tilde{r}\ll 1$,
$1/\tilde{r}^2$, is in good agreement with the numerical result. 
On the other hand, in the case of the large-$\tilde{r}$ side, because of
the very large prefactor ${\cal B}\sim 10^2$, the subleading term
remains effective. Thus, taking into account the leading term as well as
the subleading term $\sim -{\cal B}/\tilde{r}^{n+1}$,
the analytical result for $n=2.25$ matches the numerical one very well.

There are many open problems which would deserve further study.
In this paper we only discussed skyrmions in equilibrium.
Dynamical aspects, such as relaxation to the skyrmion solution,
are also important.
The dynamical aspects would be even more crucial
in possible physical realizations of our proposal.
As we have shown, the potential has to be fine-tuned to allow
a stable skyrmion solution.
In reality, it is impossible to construct a potential with
infinite precision and thus there would be some error
in the potential. This would make skyrmions absent, as
a stable solution in equilibrium.
However, we expect that, if the actual potential is close enough
to the exact one with a skyrmion solution,
the skyrmion would exist as a quasi-stable state with
a certain lifetime.
In order to discuss the feasibility of observation of a skyrmion,
we would need to estimate the lifetime of the skyrmion
in the actual potential.
We hope progress will be made on these problems, and
also on other directions related to our study.

\begin{acknowledgments}
This work was supported in part by 21st Century COE programs at Tokyo
Institute of Technology ``Nanometer-Scale Quantum Physics'' and at
Hokkaido University ``Topological Science and Technology'',
and Grant-in-Aid for Exploratory Research
No. 20654030, from MEXT, Japan.
A.T. was supported by JSPS.
I.F.H. was supported by the NSERC of Canada.

\end{acknowledgments}

\appendix
\section{Shooting method}\label{shootingmethod}
In general, iterating the discretized
time step based on the boundary condition, the equation of motion can be
numerically solved. 
Based on this observation,
we adopt the so-called shooting method in the numerical
scheme as follows:
\begin{enumerate}
\item Choose the initial velocity $d \bar{\omega}/dt(0)$
arbitrarily.
\item Given the initial velocity and the initial position
$\bar{\omega}(0)=0$, solve equation of
motion~(\ref{classical_EOM}) toward $t=\infty$.
\item If the particle goes beyond the peak of the potential
at $\bar{\omega}=\pi$, the initial velocity was too large.
\item If the particle comes back without reaching the peak
of the potential at $\bar{\omega}=\pi$, the initial velocity
was too small.
\end{enumerate}

In a true solution, the particle should approach asymptotically
the peak of the potential $\bar{\omega}=\pi$
as $t \to \infty$.
However, the asymptotic behavior in $t \to \infty$
is quite sensitive to the initial velocity, and
the numerical solution departs from the peak of the potential
in either way, depending on the tiny difference
of order of machine precision, in the initial velocity.
The sensitivity is due to the instability of the particle
at the potential peak, further enhanced by the negative
dissipation coefficient.
On the other hand, because of this sensitivity, we can
easily obtain the correct initial velocity in a high precision.

In practice, the shooting method can be
implemented as an efficient iteration
using the bisection method as follows:
\begin{enumerate}
\item Find the two values
of the initial velocity $\{ v^L_1, v^U_1 \}$,
so that $v^L_1$ is ``too small'' and $v^U_1$ is ``too large.''
\item On $n$th iteration, the correct initial velocity
should be within the range $(v^L_n, v^U_n)$.
Thus, numerically solve the equation of motion with the
midpoint initial velocity
$v^M_n \equiv (v^L_n+v^U_n)/2$.
\item If the midpoint $v^M_n$ is too small as the
initial velocity, set $v^L_{n+1} = v^M_n$,
$v^U_{n+1} = v^U_n$,
If $v^M_n$ is too large, set instead
$v^L_{n+1} = v^L_n$, $v^U_{n+1} = v^M_n$.
Go to step 2 as the $(n+1)$th iteration.
\end{enumerate}
In this way, the error in the initial velocity
decreases proportionally to $2^{-n}$ in the $n$th iteration.

\section{Asymptotic form of trap potentials}\label{appendix:asymptotics}
In order to drive the asymptotic form of trap potentials, we rewrite
equation of motion (\ref{classical_EOM}) for $t\ll -1$ and 
$t\gg 1$. Then, in this odd-dissipation case $\eta(t)$ can be
approximated as $\eta(t)\approx n$ for $t\ll -1$ and $\eta(t)\approx -n$
for $t\gg 1$. In addition, as shown by the numerical results, the
solution would stand near $\bar{\omega}(t\ll -1)\approx 0$ and 
$\bar{\omega}(t\gg 1)\approx \pi$. From these assumptions, equation
of motion~(\ref{classical_EOM}) is linearized as
\begin{equation}
 \frac{d^2\bar{\omega}}{dt^2}
 =2\bar{\omega}-n\frac{d\bar{\omega}}{dt},
\end{equation}
for $t\ll -1$, and
\begin{equation}
 \frac{d^2\bar{\omega}}{dt^2}
 =-2\pi +2\bar{\omega}+n\frac{d\bar{\omega}}{dt},
\end{equation}
for $t\gg 1$. The solutions satisfying the boundary conditions, 
$\lim_{t\to -\infty}\bar{\omega}(t)=0$ and 
$\lim_{t\to +\infty}\bar{\omega}(t)=\pi$, can be easily obtained,
respectively, as
\begin{equation}
 \bar{\omega}(t)
 =\left\{
  \begin{aligned}
   & D_1 e^{\lambda t} & &(t\ll -1) \\
   & \pi+D_2 e^{-\lambda t} & &(t\gg 1)
  \end{aligned}
  \right.,
 \label{asymp_omega}
\end{equation}
where $\lambda=\frac{-n+\sqrt{n^2+8}}{2}>0$. 
$D_1$ and $D_2$ are unknown constants. On the other hand, from 
the boundary conditions for $\eta(t)$ and the definition of the
dissipation $\eta(t)=1+\frac{d}{dt}\ln{\bar{f}_0^2}$, the asymptotic
forms of the density profile for $t\ll -1$ and $t\gg 1$ are easily
obtained as
\begin{equation}
 \bar{f}^2(t)
 \propto
  \left\{
   \begin{aligned}
    & e^{(n-1)t}  & &(t\ll -1) \\
    & e^{-(n+1)t} & &(t\gg 1)
   \end{aligned}
  \right..
 \label{asymp_f}
\end{equation}

Taking asymptotic forms~(\ref{asymp_omega}) and~(\ref{asymp_f}) from the
$t$ representation to $\tilde{r}$ representation by $t=\ln{\tilde{r}}$,
and substituting them into one of equations of motion~(\ref{ELeq1v2}),
the trap potential $\tilde{V}(\tilde{r})$ can be derived. As a result,
$\tilde{V}(\tilde{r})$ for $\tilde{r}\ll 1$ and $\tilde{r}\gg 1$ are
investigated as
\begin{equation}
 \tilde{V}(\tilde{r})
 =1-D_{1}^2\frac{\lambda^2+2}{2\tilde{r}^{2-2\lambda}}
   +\frac{n^2-1}{8\tilde{r}^2}
   -B_1\frac{\tilde{U}}{\tilde{r}^{1-n}},
 \label{asymptotic_potential1}
\end{equation}
for $\tilde{r}\ll 1$, and 
\begin{equation}
 \tilde{V}(\tilde{r})    
 =1-D_{2}^2\frac{\lambda^2+2}{2\tilde{r}^{2+2\lambda}} 
   +\frac{n^2-1}{8\tilde{r}^2}
   -B_2\frac{\tilde{U}}{\tilde{r}^{1+n}},
 \label{asymptotic_potential2}
\end{equation}
for $\tilde{r}\gg 1$. $B_{1}$ and $B_{2}$ are unknown constants
and are the prefactors of the density profile, respectively, for
$t\ll -1$ and $t\gg 1$. In particular, in the case of
$\eta(t)=-n\tanh{t}$ discussed in this paper, they turn out to be $B$
appearing in Eq.~(\ref{density_profile}): $B_1=B_2=B$.
The leading term is decided by the value of $n$. However, in
Fig.~\ref{trap_potential}, because of the very large prefactor
$B\tilde{U}\sim 10^2$, the subleading term still remains effective in
shown region of $\tilde{r}$.


\end{document}